\documentclass[english,PRE, twocolumn]{revtex4-1}
\usepackage[T1]{fontenc}
\usepackage[latin9]{inputenc}
\usepackage{amstext}
\usepackage{amssymb}
\usepackage{graphicx}
\usepackage{esint}
\usepackage{babel}
\begin{document}
\title{The Hamilton-Jacobi Equation : an intuitive approach.}
\author{Bahram Houchmandzadeh.}
\affiliation{CNRS, LIPHY, F-38000 Grenoble, France~\\
Univ. Grenoble Alpes, LIPHY, F-38000 Grenoble, France}
\begin{abstract}
The Hamilton-Jacobi equation (HJE) is one of the most elegant approach
to Lagrangian systems such as geometrical optics and classical mechanics,
establishing the duality between trajectories and waves and paving
the way naturally for the quantum mechanics. Usually, this formalism
is taught at the end of a course on analytical mechanics through its
technical aspects and its relation to canonical transformations. I
propose that the teaching of this subject be centered on this duality
along the lines proposed here, and the canonical transformations be
taught only after some familiarity with the HJE has been gained by
the students. 
\end{abstract}
\maketitle

\section{Introduction.}

There are three different formalization of classical mechanics : the
Lagrangian, the Hamiltonian and the Hamilton-Jacobi formalism. Usually,
textbooks on mechanics (see for example \citep{lanczos1986thevariational,landau1976mechanics,hand1998analytical,goldstein2011classical,calkin1645lagrangian}
) begin with the Lagrangian formalism and the variational principle,
where students discover the beauty of post-Newtonian mechanics. Historically,
this formalism was developed in analogy with optics and the principle
of Fermat\citep{lanczos1986thevariational}. Then, after a Legendre
transform, the Hamiltonian approach is introduced where students discover
the beauty of the phase space and the geometry herein. The mathematics
behind these two methods is fairly standard and more or less easily
digested by students. Finally, students come to the Hamilton-Jacobi
equation (HJE). The HJE is usually introduced after a heavy passage
through canonical transformations to uncover a first-order non-linear
partial differential equation that does not seem any more useful to
students at first glance than the former approaches. 

The aim of this short note is to make an intuitive approach to the
HJE by reversing how it is generally taught. The beauty of the HJ
approach is to uncover the duality between trajectories and wavefronts.
This duality was known in optics\citep{vohnsen2004ashort} where light
could be either investigated by rays and geometric optics (Fermat's
principle) or by wavefront (Huygens principle)\citep{crewthewave},
much before interference and the electromagnetic nature of light was
discovered. Hamilton showed that this duality can be extended to any
system described by a Lagrangian formalism, including and foremost,
mechanics. I believe that this duality and its various extensions,
specifically to quantum mechanics,are what should be taught first
and foremost to students , studied in depth. Only when the students
are familiarized with these concepts, one should introduce the canonical
transformations and the technical aspects that make this approach,
in the words of Arnold\citep{arnold2013mathematical}, ``{[}...{]}
the most powerful method known for the exact integration {[}of Hamilton
equations{]}''. At the undergraduate level, specifically to physics
students, these technical aspects seem less relevant : Arnold\citep{arnold2013mathematical}
quotes Felix Klein, who had great respect for the work of Hamilton\citep{klein1901textquotedblleftuber},
about HJ method ``\emph{that does not bring anything to the engineer
and very little to the physicist}''. Indeed, many examples of HJE
treated in the above mentioned textbooks of analytical mechanics can
be as easily treated by the Lagrangian and Hamiltonian approach.

\section{Geometrical optics.}

Eighteen century physics saw a raging debate between the particle
theory and wave theory of light\citep{euler2003lettres}. In the first
description, light is made of particles whose trajectories can be
followed and are called the ``ray paths''. In the second description,
light is made of \emph{waves}, and the ``wave front'' can be followed
exactly as we follow waves on the surface of a liquid or sounds. This
second approach was developed first by Huygens around 1680 AD\citep{crewthewave}.
In the limit of geometrical optics, when the wave length can be considered
small, these two approaches are equivalent : knowing the wave fronts,
one can deduce the ray paths and vice versa. We will detail this derivation
below, but let us first define more precisely what a wave front is
in optics.

\begin{figure}
\begin{centering}
\includegraphics[width=0.5\columnwidth]{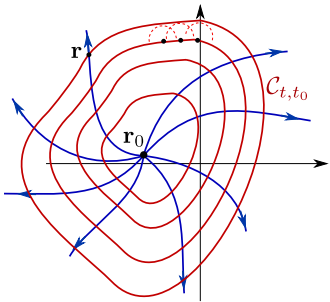}
\par\end{centering}
\caption{Wave fronts ${\cal C}$ (in red) of light emitted at point $\mathbf{r}_{0}$
at time $t_{0}$. Blue lines are the rays path. The Huygens principle
states that the wave front at time $t$ can be seen as the wave front
of light emitted at time $t-t_{\alpha}$ by the wave front at this
time (red dashed lines). \label{mar:Wave-fronts}}

\end{figure}

Consider light emitted from a point $\mathbf{r}_{0}$ at time $t_{0}$.
The boundary ${\cal C}_{t,t_{0}}$of the domain that the light has
covered at time $t$ is called the ``\emph{wave front}'' (figure
\ref{mar:Wave-fronts}) at time $t$. If the propagation medium is
homogeneous, the wave front is a sphere given by the equation 
\[
\left\Vert \mathbf{r}-\mathbf{r}_{0}\right\Vert =\frac{c}{n}(t-t_{0})
\]
where $c$ is the speed of light and $n$ the index of the propagating
medium. We can rewrite this equation as 
\[
S(\mathbf{r},t)=-(c/n)t_{0}
\]
Where the function $S(\mathbf{r},t)=\left\Vert \mathbf{r}-\mathbf{r}_{0}\right\Vert -(c/n)t$.
The relation $S(\mathbf{r},t)=-(c/n)t_{0}$ defines the collection
of points that the light (emitted at $\mathbf{r}_{0}$,$t_{0}$) has
reached at time $t$. 

We don't need to suppose that the light is emitted by a single point,
we can as well describe the wave front of the light emitted by a line
or a surface (or any at most $n-1$ dimensional object). In fact,
Huygens discovered that the wave front at time $t$ can be described
by the light emitted by the wave front at time $t-t_{\alpha}$. This
is called the Huygens principle. Finally, note that if $r_{0}\gg r$,
$\left\Vert \mathbf{r}-\mathbf{r}_{0}\right\Vert \approx r_{0}-(\mathbf{r}_{0}/r_{0}).\mathbf{r}$
and we can approximate the spherical wave by a plane one of the form
$S(\mathbf{r},t)=\mathbf{u}.\mathbf{r}-(c/n)t$ where $\mathbf{u}=-(\mathbf{r}_{0}/r_{0})$
is the direction of the plane wave propagation. 

If the medium is not homogeneous ($n=n(\mathbf{r})$ ), the wave fronts
are not spherical any more. The principle of Fermat states that the
path taken by a ray to go from a point $A$ to a point $B$ is the
one that minimizes the traveling time : 
\begin{figure}
\begin{centering}
\includegraphics[width=0.5\columnwidth]{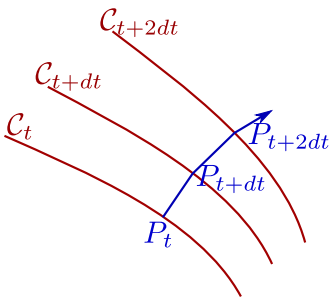}
\par\end{centering}
\caption{In geometrical optics in isotropic media, trajectories $P_{t}$ of
the light rays and wave fronts are orthogonal. Therefore, trajectories
can be recovered from the wave front: from the point $P_{t}$ on the
wave front ${\cal C}_{t}$, draw the orthogonal to the wave front
and recover the point $P_{t+dt}$ at which it intercepts the wave
front ${\cal C}_{t+dt}$. Proceeds by recurrence. \label{mar:trajectories--waves}}
\end{figure}
\[
T=\frac{1}{c}\int_{A}^{B}nds
\]
where $ds$ is the element of arc length along a path. In order to
compute a wave front now, one has to compute the ray paths and collect
points along the path that have been reached at a given time $t$.
If the medium is isotropic (\emph{i.e. }not like a crystal with particular
directions of propagation), it can be shown that ray paths and wave
fronts are orthogonal (see below). In this case, deducing the wave
fronts from the ray paths is simple. On the other hand, \emph{if}
we knew the wave fronts, we could compute the ray paths (figure \ref{mar:trajectories--waves}).
Paths and wave fronts are dual objects linked together through an
orthogonality. 

Even if the medium is \emph{not} isotropic, we can still compute the
wave front from the rays, and vice versa. All we need is a relation
between the tangent to the ray path (let's call it $\dot{\mathbf{q}}$)
at a point and the normal to the wave front (call it $\mathbf{p}$)
at the same point. We will come to this subject in more general detail
in the next sections. 

\section{Basic notions of analytical mechanics.}

Very soon after the publication of \emph{Principia }by Newton (1684),
Bernoulli challenged (1696) the scientific community to find the fastest
path that, under gravity, brings a mass from point $A$ to point $B$.
The analogy with optics and the Fermat's principle was not lost on
the mathematicians who responded to the challenge\citep{erlichson1999johannbernoullitextbackslashtextquotesingles}.
This analogy was then fully developed in subsequent years \citep{lagrange2000mecanique}
and took its definitive form under the name of Euler-Lagrange equation. 

\begin{figure}
\begin{centering}
\includegraphics[width=0.5\columnwidth]{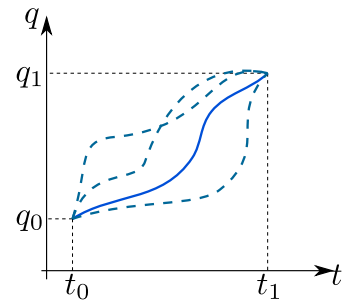}
\par\end{centering}
\caption{The trajectory chosen by an object (solid line) optimizes the action
compared to all other possible trajectories (dashed lines) (\ref{eq:action-def})
\label{mar:The-trajectory-chosen}}

\end{figure}

The foundation of analytical mechanics is based on a variational principles:
Given a Lagrangian ${\cal L}(\dot{q},q,t)$, an object (be it a particle
or a ray of light) \emph{chooses} the trajectory $q(t)$ that makes
the \emph{action} 
\begin{equation}
S=\int_{t_{0},q_{0}}^{t_{1},q_{1}}{\cal L}(\dot{q},q,t)dt\label{eq:action-def}
\end{equation}
stationary (figure \ref{mar:The-trajectory-chosen}). The action depends
on the \emph{end points} $(t_{0},q_{0})$ and $(t_{1},q_{1})$ and
the trajectory $q(t)$ must obey the Euler-Lagrange equation
\begin{equation}
\frac{d}{dt}\frac{\partial{\cal L}}{\partial\dot{q}}-\frac{\partial{\cal L}}{\partial q}=0\label{eq:euler-lagrange}
\end{equation}
For a classical particle, the Lagrangian is the difference between
the kinetic and the potential energy ${\cal L}=T-V$, while for geometrical
optics, the Lagrangian is the traveling time. 

We can reformulate equation (\ref{eq:euler-lagrange}) by making a
Legendre transform. Defining the momentum 
\begin{equation}
p=\frac{\partial{\cal L}}{\partial\dot{q}}\label{eq:momentum}
\end{equation}
expressing $\dot{q}$ as a function of $p$ and defining $H(p,q,t)=p\dot{q}-{\cal L}$,
we obtain the Hamilton equations 
\begin{equation}
\frac{dq}{dt}=\frac{\partial H}{\partial p}\,\,\,;\,\,\,\frac{dp}{dt}=-\frac{\partial H}{\partial q}\label{eq:hamiltone:phasespace}
\end{equation}
which allows us to move to the phase space and have a more geometrical
view of the trajectories. One consequence of the above equation is
the variation of $H$ as a function of time along a trajectory:
\begin{equation}
dH=\frac{\partial H}{\partial p}dp+\frac{\partial H}{\partial q}dq+\frac{\partial H}{\partial t}dt=\frac{\partial H}{\partial t}dt\label{eq:hamilton}
\end{equation}
Therefore, if the Hamiltonian does not depend explicitly on time,
the Hamiltonian is conserved along a trajectory: $H=E$.

\begin{figure}
\begin{centering}
\includegraphics[width=0.6\columnwidth]{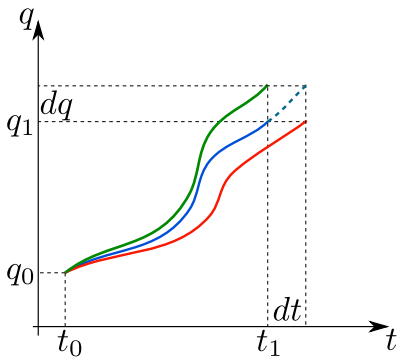}
\par\end{centering}
\caption{Varying the end points of a movement.\label{mar:Varying-the-end}}

\end{figure}

In the above two formulation of analytical mechanics, the action $S()$
itself plays little explicit role; what is important is the differential
equations (\ref{eq:euler-lagrange}) or (\ref{eq:hamiltone:phasespace})
whose solution determines the trajectory. However, Let us have a closer
look at the action itself. By action $S$ here we mean the integral
expression (\ref{eq:action-def}) when the particle moves along the
\emph{optimal} path. Even though the absolute value of $S$ can be
hard to compute analytically, we can compute its \emph{variation}
if we vary the end points (figure \ref{mar:Varying-the-end}). We
will keep here the initial point fixed and vary the final end point
either by $dt$ or $dq$. 

We begin by keeping the final time fixed at $t_{1}$ but move the
final position by $dq$ (figure \ref{mar:Varying-the-end}). The trajectory
$q(t)$ will vary by $\delta q(t)$ where $\delta q(t_{0})=0$ and
$\delta q(t_{1})=dq$. The variation in $S$ is 
\begin{equation}
\delta S=\int_{t_{0}}^{t_{1}}\left\{ \frac{\partial{\cal L}}{\partial\dot{q}}\delta\dot{q}+\frac{\partial{\cal L}}{\partial q}\delta q\right\} dt\label{eq:deltaS}
\end{equation}
However, the trajectories obey the Euler-Lagrange equation (\ref{eq:euler-lagrange})
and we must have 
\[
\frac{\partial{\cal L}}{\partial q}=\frac{d}{dt}\frac{\partial{\cal L}}{\partial\dot{q}}
\]
On the other hand, $\delta\dot{q}=d(\delta q)/dt$. Using these relations,
we can rewrite equation (\ref{eq:deltaS}) as 
\begin{eqnarray*}
\delta S & = & \int_{t_{0}}^{t_{1}}\left\{ \frac{\partial{\cal L}}{\partial\dot{q}}\frac{d\left(\delta q\right)}{dt}+\frac{d}{dt}\left(\frac{\partial{\cal L}}{\partial\dot{q}}\right)\delta q\right\} dt\\
 & = & \int_{t_{0}}^{t_{1}}\frac{d}{dt}\left\{ \frac{\partial{\cal L}}{\partial\dot{q}}\delta q\right\} dt\\
 & = & \left[\frac{\partial{\cal L}}{\partial\dot{q}}\delta q\right]_{t_{0}}^{t_{1}}=\left.\frac{\partial{\cal L}}{\partial\dot{q}}\right|_{t_{1}}dq
\end{eqnarray*}
As we have kept the final time fixed, $\delta S=\left(\partial S/\partial q\right)dq$
and therefore 
\begin{equation}
\frac{\partial S}{\partial q}=\left.\frac{\partial{\cal L}}{\partial\dot{q}}\right|_{t_{1}}=p(t_{1})\label{eq:HJ1}
\end{equation}
If we vary the end point $q_{1}$, the relative variation in $S$
is the\emph{ momentum} $p$ at the end point. 

To compute the variation of $S$ as a function of the end point's
time, consider letting the original trajectory to continue along its
optimal path. Then $dS={\cal L}dt.$ On the other hand 
\[
dS={\cal L}dt=\frac{\partial S}{\partial q}dq+\frac{\partial S}{\partial t}dt
\]
Using our previous result (\ref{eq:HJ1}), we have 
\[
{\cal L}dt=pdq+\frac{\partial S}{\partial t}dt=\left(p\dot{q}+\frac{\partial S}{\partial t}\right)dt
\]
and therefore 
\begin{equation}
\frac{\partial S}{\partial t}={\cal L}-p\dot{q}=-H\label{eq:HJ2}
\end{equation}
Relation (\ref{eq:HJ1},\ref{eq:HJ2}) are very general results of
variational calculus with varying end points and are not restricted
to mechanics. The contact angle of a liquid droplet on a solid surface
is obtained for example by these computations. Note also that even
though we derived these equations in one dimension of space, they
are trivially generalized to any dimension. 

\section{General wave fronts.}

In geometrical optics, we had used the traveling time to define the
wave front. But the traveling time is just one example of action and
variational principles. In analogy with optics, let us define the
function $S_{\mathbf{q}_{0},t_{0}}(\mathbf{q},t)$ as the action of
a particle that arrives at $(\mathbf{q},t)$ after leaving $(\mathbf{q}_{0},t_{0})$,
following its optimal path. By this function, we can associate to
each point $(\mathbf{q},t)$ a value in space-time. Then, $S(\mathbf{q},t)=C$
defines an $n-1$ dimensional surface ${\cal C}_{t}$, \emph{i.e.
}the collection of points $\mathbf{q}$ that have the same value $C$
of action at time $t$. Figure \ref{mar:Wave-fronts} that illustrated
wave front in optics illustrates similarly the general wavefronts
of action. 

Consider for example a classical free particle, whose trajectories
are straight lines with constant speed $v=\text{\ensuremath{\left\Vert \mathbf{q}-\mathbf{q}_{0}\right\Vert }}/(t-t_{0})$.
The action is therefore 
\[
S(\mathbf{q},t)=\frac{m}{2}v^{2}(t-t_{0})=\frac{m}{2}\text{\ensuremath{\left\Vert \mathbf{q}-\mathbf{q}_{0}\right\Vert }}^{2}/(t-t_{0})
\]
and the curves ${\cal C}_{t}$ are spheres of radius proportional
to $\sqrt{2(t-t_{0})/m}$. If the initial point is far away from the
region of interest ($\left|t\right|\ll\left|t_{0}\right|$, $q\ll q_{0}$),
we can develop the above expression and write it, to the first order
in $q$,$t$ :
\begin{eqnarray}
S(\mathbf{q},t) & \approx & \frac{m}{2t_{0}}\left(q_{0}^{2}-2\mathbf{q}_{0}.\mathbf{q}\right)(-1-t/t_{0})\nonumber \\
 & = & S_{0}+\mathbf{p}.\mathbf{q}-Et\label{eq:action-wave}
\end{eqnarray}
where we have defined the constants $\mathbf{p}=m\mathbf{q}_{0}/t_{0}$
and $E=(1/2)mq_{0}^{2}/t_{0}^{2}$. In this case, the action is a
plane wave. 

\begin{figure}
\begin{centering}
\includegraphics[width=0.5\columnwidth]{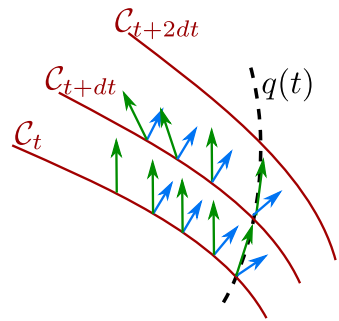}
\par\end{centering}
\caption{From known wave fronts ${\cal C}_{t}$(in red) to trajectories : at
each point, the normal to the wave front $\mathbf{p}=\partial S/\partial\mathbf{q}$
(in blue) can be computed ; knowing $\mathbf{p}$, we can compute
the tangent to the trajectory $\dot{\mathbf{q}}$ ( in green ) and
find a trajectory following a given line of tangents. The procedure
is trivially generalized to higher dimensional space where $\mathbf{q}$
collects the coordinates of many particles. \label{mar:wave-trajectories-general}}

\end{figure}

We have defined the wave front as the collection of points $\mathbf{q}$
at time $t$ for which $S(\mathbf{q},t)=\text{const.}$ To compute
the wavefronts however, we have relied on the knowledge of trajectories.
To go further, we need to derive an independent equation from which
$S()$ can be computed directly, without any a priori knowledge of
trajectories. For this purpose, we just have to recall from the last
section (\ref{eq:HJ1},\ref{eq:HJ2}) that we can compute the \emph{variation}
of $S$ as a function of the variation of its end points:

\begin{equation}
\frac{\partial S}{\partial\mathbf{q}}=\mathbf{p}\,\,\,\,;\,\,\,\,\frac{\partial S}{\partial t}=-H\label{eq:closedform}
\end{equation}
where $\partial S/\partial\mathbf{q}=(\partial_{q_{1}}S,\partial_{q_{2}}S,...)$.
Note that this a generalization of the free particle case where (according
to \ref{eq:action-wave}), $dS=\mathbf{p}d\mathbf{q}-Edt$. Now, we
know that $H=H(\mathbf{q},\mathbf{p},t)$, therefore combining the
above two expressions, we have 

\begin{equation}
\frac{\partial S}{\partial t}+H\left(\mathbf{q},\frac{\partial S}{\partial\mathbf{q}},t\right)=0\label{eq:HJE}
\end{equation}
which is a \emph{first order} PDE and called the Hamilton-Jacobi equation
(HJE). If we can solve this equation and find the wave fronts, then
we can deduce the trajectories from the wavefronts. The procedure
is similar to what we did in geometrical optics : At each time $t$,
we know the wave front $S$, and therefore, we can compute the momentum
at points $\mathbf{q}$: $\mathbf{p}(\mathbf{q})=\partial S/\partial\mathbf{q}$
(figure \ref{mar:wave-trajectories-general}). This vector is related
to the tangent to a trajectory $\dot{\mathbf{q}}$ through the relation
\[
\mathbf{p}=\frac{\partial{\cal L}}{\partial\dot{\mathbf{q}}}
\]
By resolving the above relation, we can compute $\dot{\mathbf{q}}$
at each point of space at each time : 
\begin{equation}
\dot{\mathbf{q}}=f(\mathbf{q},t)\label{eq:firstorder}
\end{equation}
If we knew the wave fronts, the second order differential equations
of Euler-Lagrange (equation \ref{eq:euler-lagrange}) are transformed
into ordinary first order differential equations (\ref{eq:firstorder})
as above. For the simplest mechanical systems with one particle and
a potential $V(\mathbf{q},t),$ $\mathbf{p}$ and $\mathbf{\dot{q}}$
are co-linear and the construction is really similar to optics. 

We can further simplify the HJE (eq. \ref{eq:HJE}) if the function
$H$ does not contain $t$ explicitly. In this case, we can \emph{separate}
the function $S$ into 
\begin{equation}
S(\mathbf{q},t)=W(\mathbf{q})-Et\label{eq:separable-S}
\end{equation}
where the function $W()$ (often called Hamilton principal function)
obeys the relation 
\begin{figure}
\begin{centering}
\includegraphics[width=0.6\columnwidth]{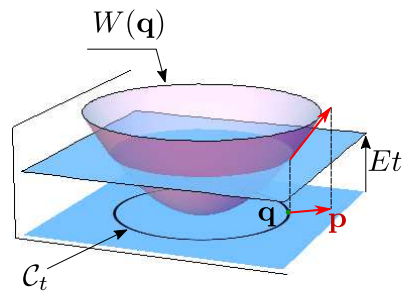}
\par\end{centering}
\caption{An illustration of the wave front in a two dimensional space where
the function $W(\mathbf{q})$ is represented as a surface in three
dimension. The wave front ${\cal C}_{S}$ is the contour plot of the
function $W(\mathbf{q})$. At any given point $\mathbf{q},$the momentum
is given by $\mathbf{p}=\nabla W$\label{mar:cut}}
\end{figure}
\[
H\left(\mathbf{q},\frac{\partial W}{\partial\mathbf{q}}\right)=E
\]
Once $W()$ is solved for, we can find the wave fronts by slicing
the function $W()$ at different ``heights'' : at a given time $t$,
we collects all points $\mathbf{q}$ such that $W(\mathbf{q})=Et+\text{const.}$
into the wave front ${\cal C}_{t}$ (figure \ref{mar:cut}).

\section{Examples.}

\subsection{One particle.}

\begin{figure}
\begin{centering}
\includegraphics[width=0.5\columnwidth]{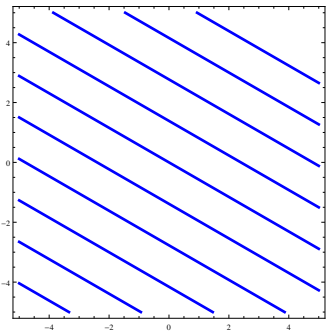}
\par\end{centering}
\caption{Contour plot of $W(q_{1},q_{2})$ of free particle in 2 dimensions
(relation \ref{eq:freepartW}) for $u_{1}=\cos\theta=1/2$. \label{mar:Contour-freepart} }

\end{figure}

Consider one classical free particle with the Lagrangian ${\cal {\cal L}}=(1/2)m\dot{\mathbf{q}}^{2}$,
$\mathbf{p}=m\dot{\mathbf{q}}$ and $H=\mathbf{p}^{2}/2m$ where we
use the square of a vector as a shorthand: $\mathbf{u}^{2}=\mathbf{u}.\mathbf{u}$.
Therefore, the HJE is simply 
\begin{equation}
\frac{\partial S}{\partial t}+\frac{1}{2m}\left(\frac{\partial S}{\partial\mathbf{q}}\right)^{2}=0\label{eq:HJE:freeparticle}
\end{equation}
It is straightforward to check that the spherical wave $S=m(\mathbf{q}-\mathbf{q}_{0})^{2}/2(t-t_{0})$
is a solution of the above equation, where $\mathbf{q}_{0}$ and $t_{0}$
are some constants. We can also look for a \emph{separable} solution
of the form $S=W-Et$, in which case
\[
\frac{1}{2m}\left(\frac{\partial W}{\partial\mathbf{q}}\right)^{2}=E
\]
To solve this PDE, we can search for further separability in the form
of 
\begin{equation}
W(\mathbf{q})=\sqrt{2m}\sum_{i}w_{i}(q_{i})\label{eq:variable-sep}
\end{equation}
and solve the equations $dw_{i}/dq_{i}=\sqrt{e_{i}}$ where $e_{i}$
are integration constants. The solution of these equations are $w_{i}(q_{i})=\sqrt{e_{i}}q_{i}+C_{i}$
with the constraints $\sum_{i}e_{i}=E$ and $C_{i}$ another set of
integration constants. The complete solution is then a plane wave
with (figure \ref{mar:Contour-freepart})
\begin{equation}
W(\mathbf{q})=\sqrt{2mE}\sum_{i}u_{i}q_{i}+C'_{i}\label{eq:freepartW}
\end{equation}
where $u_{i}=\sqrt{e_{i}/E}$ are the integration constants. We collect
the constants $u_{i}$ into a constant vector $\mathbf{v}$ such that
$v_{i}=vu_{i}$, $E=(1/2)mv^{2}$ and write (figure \ref{mar:Contour-freepart})
\[
W(\mathbf{q})=m\mathbf{v}.\mathbf{q}+C
\]
 
Now that we know the wave front, if we wish so, we can deduce the
trajectories : the moment is given by $\mathbf{p}=\partial W/\partial\mathbf{q}=m\mathbf{v}$.
From the Lagrangian, we know that $\dot{\mathbf{q}}=\mathbf{p}/m$,
and therefore $\dot{\mathbf{q}}=\mathbf{v}$ and $\mathbf{q}=\mathbf{v}t+\mathbf{q}_{0}$
where $\mathbf{q}_{0}$ is another integration constant. 

For a classical free particle, the HJE is obviously an overkill. The
purpose of this example is to illustrate how the solution of the HJE
with the integration constant $\mathbf{v}$ leads to the trajectories.
It is straightforward to check that the spherical wave solution leads
to the same result for trajectories.

Adding a potential $V(\mathbf{q})$ to the problem give rise to the
HJE 
\begin{equation}
\frac{\partial S}{\partial t}+\frac{1}{2m}\left(\frac{\partial S}{\partial\mathbf{q}}\right)^{2}=-V(\mathbf{q})\label{eq:classical:parabolic}
\end{equation}
There exist a systematic method to search for the solution of this
equation, called \emph{canonical transformations} (see for example
\citep[section 10.4]{goldstein2011classical}). If however, the potential
is itself separable $V(\mathbf{q})=\sum_{i}V_{i}(q_{i})$, we can
look for a separable solution of the HJE as before. As an illustration,
consider the simple one dimensional harmonic oscillator with $V(q)=(1/2)kq^{2}$.
Extension to higher dimensional case is trivial but harder to present
graphically. Setting $S=W-Et$ we have 
\[
\frac{dW}{dq}=\sqrt{2mE}\sqrt{1-\frac{x^{2}}{\ell^{2}}}
\]
where $\ell^{2}=2E/k$. Setting $q=\ell\sin\theta$ transforms the
above equation into 
\begin{figure}
\begin{centering}
\includegraphics[width=0.6\columnwidth]{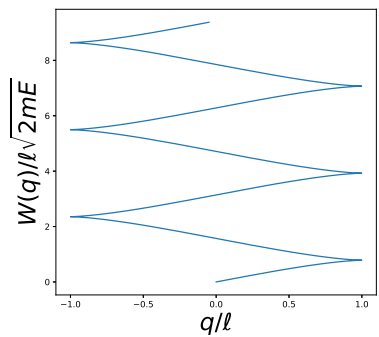}
\par\end{centering}
\caption{The Hamilton principal function $W(q)$ for the uni-dimensional harmonic
oscillator.\label{mar:harmonic:osc}}

\end{figure}
\[
\frac{dW}{d\theta}=\ell\sqrt{2mE}\cos^{2}\theta
\]
that integrates directly
\[
W(\theta)=\frac{1}{2}\ell\sqrt{2mE}(\theta+\frac{1}{2}\sin2\theta)+C
\]
Figure \ref{mar:harmonic:osc} displays a plot of $W(q)$ as a function
of $q$. It can be observed that the function $W()$ is multivalued
and at its ``turning points'', $p=\partial W/\partial q=0$, a fact
that is common to all bounded mechanical systems. 

\subsection{Relativistic particle.}

We distinguish here explicitly between time and space coordinate for
more clarity at the expense of elegance. Consider a free relativistic
particle whose action is given by its Minkowski arc length 
\[
S=-m\int_{A}^{B}ds
\]
where (in natural units $c=1$ ) ${\cal L}dt=$$-mds=-m\sqrt{dt^{2}-d\mathbf{x}d\mathbf{x}}=-m\sqrt{1-\dot{\mathbf{x}}^{2}}dt$.
We have 
\[
\mathbf{p}=\frac{\partial{\cal L}}{\partial\dot{\mathbf{x}}}=\frac{m\dot{\mathbf{x}}}{\sqrt{1-\dot{\mathbf{x}}^{2}}}
\]
and therefore 
\[
H=\mathbf{p}\dot{\mathbf{x}}-{\cal L}=\sqrt{m^{2}+\mathbf{p}^{2}}
\]
The HJ equation is therefore 
\[
\left(\frac{\partial S}{\partial t}\right)=-\sqrt{m^{2}+\left(\frac{\partial S}{\partial\mathbf{x}}\right)^{2}}
\]
or 
\begin{equation}
\left(\frac{\partial S}{\partial t}\right)^{2}-\left(\frac{\partial S}{\partial\mathbf{x}}\right)^{2}=m^{2}\label{eq:HJE:relativistic}
\end{equation}
Note that the parabolic PDE of a classical dynamics becomes a wave
equation when we consider the relativistic dynamics. This is exactly
how the Schrodinger equation transforms into the Klein-Gordon one,
\emph{i.e.} the relativistic wave equation for spineless particles.
This can be extended to the case of a particle with in an electromagnetic
field by considering
\begin{eqnarray*}
{\cal L}dt & = & -mds-qd\mathbf{s}.\mathbf{A}
\end{eqnarray*}
where the four vector $\mathbf{A}=(-\phi,\vec{A})$, $\phi$ is the
electromagnetic potential and $\vec{A}$ the (three) vector potential. 

\subsection{Geometrical optics.}

Consider light propagating in an isotropic medium. The action is the
total traveling time
\[
S=\int_{A}^{B}nds
\]
where $n(\mathbf{q})$ is the index of the medium at position $\mathbf{q}$,
$ds$ is the arc length along a trajectory and we have set the speed
of light in vacuum $c=1$. This is called the principle of Fermat.
For simplicity, we will consider a two-dimensional medium where $x$
is used as the integration variable and $ds=\sqrt{dx^{2}+dy^{2}}=\sqrt{1+y'^{2}}dx$
; the Lagrangian is 
\[
{\cal L}=n(x,y)\sqrt{1+y'^{2}}
\]
and by definition, 
\begin{equation}
p=\frac{\partial{\cal L}}{\partial y'}=n\frac{y'}{\sqrt{1+y'^{2}}}\label{eq:momentum:geom}
\end{equation}
if we set $\theta$ as the angle between the tangent to the trajectory
and the $x$ axis, the above relation is simply $p=n\sin\theta$,
which is the conserved quantity if $n=n(x)$ (Snell's law). Solving
relation \ref{eq:momentum:geom} in $y'$, we have $y'=p/\sqrt{n^{2}-p^{2}}$
and therefore the Hamiltonian is 
\[
H=py'-{\cal L}=-\sqrt{n^{2}-p^{2}}
\]
The HJE is then 
\[
\left(\frac{\partial S}{\partial x}\right)-\sqrt{n^{2}-\left(\frac{\partial S}{\partial y}\right)^{2}}=0
\]
or in other words, 
\begin{equation}
\left(\frac{\partial S}{\partial x}\right)^{2}+\left(\frac{\partial S}{\partial y}\right)^{2}=n^{2}\label{eq:eikonal}
\end{equation}
The above expression, called the eikonal equation, is the fundamental
equation of geometrical optics. In the Hamilton-Jacobi approach, its
resemblance to relativistic particle is obvious. We will see below
that the eikonal equation can be obtained through approximation of
the wave equation. 

\section{Waves and particles.}

For about 50 years after its introduction, the Hamilton-Jacobi equation
was considered a beautiful but useless tool. With the advent of quantum
mechanics, Schrodinger realized that this equation is the natural
road to formulating a ``wave'' equation for particles. The approach
was as follow : geometrical optic is an approximation of the Maxwell
equations that neglects interference effect. We know the Maxwell equation
and the approximation procedure to get to geometrical optics. Schrodinger
realized that classical mechanics can be such an approximation of
a more complicated theory and reverse engineered the geometrical optics
approximation to get to his famous equation in 1926. The detail of
this procedure and its connection to Hamilton-Jacobi equation is beautifully
written by Massoliver and Ros\citep{masoliver2009fromclassical} and
we don't develop it here. However, it is very simple to show that
classical mechanics is an approximation of the quantum mechanics. 

Consider the Schrodinger equation 
\[
i\hbar\frac{\partial\psi}{\partial t}=-\frac{\hbar^{2}}{2m}\frac{\partial^{2}\psi}{\partial x^{2}}+V(x)\psi
\]
using a standard change of function 
\begin{equation}
\psi=e^{iS/\hbar}\label{eq:transpsi}
\end{equation}
the Schrodinger equation transforms into 
\begin{equation}
-\frac{\partial S}{\partial t}=-\frac{i\hbar}{2m}\frac{\partial^{2}S}{\partial x^{2}}+\frac{1}{2m}\left(\frac{\partial S}{\partial x}\right)^{2}+V(x)\label{eq:schrodinger}
\end{equation}
we see that the above equation, when we neglect the term in $\hbar$,
reduces exactly to the classical HJE (\ref{eq:classical:parabolic}):
the classical mechanics is indeed the limit of quantum mechanics when
$\hbar\rightarrow0$. 

The transformation (\ref{eq:transpsi}), called the ansatz of Sommerfield
and Runge\citep{sommerfeld_anwendung_1911}, was nothing unusual at
the time of Schrodinger and is used to recover the geometrical optics
from the wave equation ( see\citep{cornbleet1983geometrical} for
a review). Consider the equation of an electromagnetic wave propagating
through space, where the index of refraction is not supposed to be
constant : 
\begin{equation}
\frac{\partial^{2}\psi}{\partial t^{2}}=v^{2}\nabla^{2}\psi\label{eq:wave}
\end{equation}
where $\psi$ is any component of the electromagnetic tensor or the
vector potential and $v=c/n$ where $c$ is the speed of light and
$n$ the index of the medium. We look for a solution of the form
\begin{equation}
\psi(t)=A(\mathbf{r})\exp\left(ik_{0}\left(\phi(\mathbf{r})-ct\right)\right)\label{eq:erzatz}
\end{equation}
in analogy with plane waves when $n=\text{const}.$ $k_{0}=2\pi/\lambda_{0}$
is the wave number and $\lambda_{0}$ is the wave length in vacuum
; $A$ (the amplitude ) and $\phi$ (the phase) are real functions.
Note that the total phase 
\[
\Phi(\mathbf{r})=\phi(\mathbf{r})-ct
\]
has the same structure as the function $S$ in relation (\ref{eq:separable-S})
and $\phi()$ plays the same role as the function $W()$. 

Plugging expression (\ref{eq:erzatz}) into (\ref{eq:wave}), separating
the real and the imaginary part, we have: 
\begin{eqnarray}
\nabla^{2}A-Ak_{0}^{2}\left(\nabla\phi\right)^{2} & = & -k_{0}^{2}n^{2}A\label{eq:wavereal}\\
2\left(\nabla\phi\right)\left(\nabla A\right)+A\nabla^{2}\phi & = & 0\label{eq:waveimagine}
\end{eqnarray}
The geometrical optics is obtained from the wave equation by letting
$\lambda_{0}\rightarrow0$, \emph{i.e. }when we assume that the scale
of variation in the index is large compared to the wave length, or
equivalently, when $\left|\nabla^{2}A/A\right|\ll k_{0}^{2}$. Neglecting
the $\nabla^{2}A$ term is relation (\ref{eq:wavereal}), we obtain
an equation for the phase $\phi$ alone:
\begin{equation}
\left(\nabla\phi\right)^{2}=n^{2}\label{eq:eikonal2}
\end{equation}
which is the eikonal equation we had already obtained from the principle
of Fermat (eq. \ref{eq:eikonal}).

\section{Conclusion.}

The Hamilton-Jacobi equation is one of the most elegant and beautiful
approach to mechanics with far reaching consequences in many adjacent
fields such as quantum mechanics and probability theory. Unfortunately,
its beauty is lost to many students learning the basics of analytical
mechanics. An informal and statistically non-significant inquiry of
practicing physicists suggests that even among scientists, Hamilton-Jacobi
brings up mostly (if any) memories of arcane transformations with
no observable use. 

The materials developed in this short article, which does not contain
the usual mathematical complexity found in most textbooks, can be
covered in one or two lectures and I hope help students to get a basic
understanding of the Hamilton-Jacobi approach to variational systems. 

\paragraph{Acknowledgment. }

I'm grateful to Marcel Vallade for detailed reading of the manuscript
and fruitful discussions.

\bibliographystyle{unsrt}

\begin{thebibliography}{}

\end{thebibliography}


\begin{thebibliography}{10}

\bibitem{lanczos1986thevariational}
Cornelius Lanczos.
\newblock {\em The {Variational} {Principles} of {Mechanics}}.
\newblock Dover Publications, New York, 4th revised ed. edition edition, March
  1986.

\bibitem{landau1976mechanics}
L.~D. Landau and E.~M. Lifshitz.
\newblock {\em Mechanics: {Volume} 1}.
\newblock Butterworth-Heinemann, Amsterdam u.a, 3 edition, January 1976.

\bibitem{hand1998analytical}
Louis~N. Hand and Janet~D. Finch.
\newblock {\em Analytical {Mechanics}}.
\newblock Cambridge University Press, 1 edition, November 1998.

\bibitem{goldstein2011classical}
Herbert Goldstein and Charles P. Poole \&~John Safko.
\newblock {\em {CLASSICAL} {MECHANICS}}.
\newblock Pearson Education, January 2011.

\bibitem{calkin1645lagrangian}
M.~G. Calkin.
\newblock {\em Lagrangian and {Hamiltonian} {Mechanics} by {M}. {G}.
  {Calkin}(1996-07-04)}.
\newblock World Scientific Publishing Company, 1645.

\bibitem{vohnsen2004ashort}
Brian Vohnsen.
\newblock A {Short} {History} of {Optics}.
\newblock {\em Physica Scripta}, 2004(T109):75, 2004.

\bibitem{crewthewave}
Henry Crew.
\newblock {\em The wave theory of light; memoirs of {Huygens}, {Young} and
  {Fresnel}}.

\bibitem{arnold2013mathematical}
V.~I. Arnol'd, K.~Vogtmann, and A.~Weinstein.
\newblock {\em Mathematical {Methods} of {Classical} {Mechanics}}.
\newblock Springer, 2 edition, 2013.

\bibitem{klein1901textquotedblleftuber}
F.~Klein.
\newblock {\textquotedblleft}{\"u}ber das {Brunssche}
  {Eikonal}{\textquotedblright}.
\newblock {\em Zeitscrift f. Mathematik u. Physik}, 46, 1901 [Translation by D.
  H. Delphenich, http://www.neo-classical-physics.info].

\bibitem{euler2003lettres}
Leonhard Euler.
\newblock {\em Lettres {\`a} une princesse d'{Allemagne}}.
\newblock Presses Polytechniques et Universitaires Romandes, Lausanne, 2003.

\bibitem{erlichson1999johannbernoullitextbackslashtextquotesingles}
Herman Erlichson.
\newblock Johann {Bernoulli}{\textbackslash}textquotesingles brachistochrone
  solution using {Fermat}{\textbackslash}textquotesingles principle of least
  time.
\newblock {\em European Journal of Physics}, 20(5):299--304, July 1999.

\bibitem{lagrange2000mecanique}
Joseph-Louis Lagrange.
\newblock {\em M{\'e}canique analytique}.
\newblock Jacques Gabay, Paris, 2000.

\bibitem{masoliver2009fromclassical}
Jaume Masoliver and Ana Ros.
\newblock From classical to quantum mechanics through optics.
\newblock {\em European Journal of Physics}, 31(1):171--192, November 2009.

\bibitem{sommerfeld_anwendung_1911}
A~Sommerfeld and J.~Runge.
\newblock Anwendung der {Vektorrechnung} auf die {Grundlagen} der
  {Geometrischen} {Optik}.
\newblock {\em Ann. Phys. (Leipzig)}, 35:277--298, 1911 [Translation by D. H.
  Delphenich, http://www.neo-classical-physics.info].

\bibitem{cornbleet1983geometrical}
S.~Cornbleet.
\newblock Geometrical optics reviewed: {A} new light on an old subject.
\newblock {\em Proceedings of the IEEE}, 71(4):471--502, April 1983.

\end{thebibliography}

\end{document}